\documentclass[12pt]{iopart}

\usepackage{xcolor}
\usepackage{graphicx}
\usepackage{subcaption}
\usepackage{hyperref}
\usepackage{cite}
\usepackage{booktabs}
\usepackage{lineno}
\usepackage{enumitem}

\begin{document}

\title[]{Learning digital signal processing using an interactive Jupyter notebook and smartphone accelerometer data}

\author{P. Pirinen$^1$, P. Klein$^2$, S. Z. Lahme$^2$, A. Lehtinen$^{1,3}$, L. Ron\v cevi\' c$^4$, and A. Susac$^4$}
\address{$^1$Department of Physics, P.O. Box 35, 40014 University of Jyväskylä, Finland}
\address{$^2$Faculty of Physics, Physics Education Research, University of Göttingen, Friedrich-Hund-Platz 1, 37077, Göttingen, Germany}
\address{$^3$Department of Teacher Education, P.O. Box 35, 40014 University of Jyväskylä, Finland}
\address{$^4$Department of Applied Physics, Faculty of Electrical Engineering and Computing, University of Zagreb, Unska 3, 10000, Zagreb, Croatia}
\ead{pekka.a.pirinen@gmail.com}
\vspace{10pt}

\begin{abstract}
Digital signal processing is a valuable practical skill for the contemporary physicist, yet in physics curricula its central concepts are often introduced either in method courses in a highly abstract and mathematics-oriented manner or in lab work with little explicit attention. In this paper, we present an experimental task in which we focus on a practical implementation of the discrete Fourier transform (DFT) in an everyday context of vibration analysis using data collected by a smartphone accelerometer. Students are accompanied in the experiment by a Jupyter notebook companion, which serves as an interactive instruction sheet and a tool for data analysis. The task is suitable for beyond-first-year university physics students with some prior experience in uncertainty analysis, data representation, and data analysis. Based on our observations the experiment is very engaging. Students have consistently reported interest in the experiment and they have found it a good demonstration of the DFT method. 
\end{abstract}

\section{Introduction} \label{sec:intro}

Digital signal processing is an important skill for physicists, with a multitude of uses in experimental physics (see, for example, Ref. \cite{abbott2020} for signal processing in the famous LIGO-Virgo gravitational wave experiments) and industry (for example detecting faults in machinery via vibration analysis \cite{muniz2023}). Some basic elements of digital signal processing are typically included implicitly in undergraduate laboratory courses, but learning objectives related to concepts and methods of signal processing are often hidden behind more explicit objectives related to reinforcing physics concepts. 

We present an experimental task in which students analyze vibrations by performing a discrete Fourier transform (DFT) on data measured by a smartphone accelerometer. Students are accompanied in their investigation by a Jupyter notebook, which first introduces the concepts and methods relevant to the experiment and provides Python scripts for analyzing the collected data. Students ultimately measure the frequency components of an assumed periodic signal found at home that they want to investigate, either by picking one from our suggestions (for example the spin-dry rotation frequency of a washing machine or one's own heart rate) or coming up with their own target for investigation. The task can be used to teach the basics of digital signal processing in an everyday physics context to beyond-first-year university physics students, and it is perfectly suitable for a distance-learning setting although it can be conducted on campus as well.

Physics lab tasks utilizing the DFT and its algorithm implementation fast Fourier transform (FFT) have existed for decades \cite{matthys1982,lambert1985}. Ref. \cite{kraftmakher2012} showcases several examples with electric circuits. Ref. \cite{weigman1993} uses the FFT in a mechanics context to analyze the eigenfrequencies of coupled harmonic oscillators from force transducer data. In Ref. \cite{hammer2011} a similar vibration analysis as in the present work was introduced utilizing computer-assisted measurements and ready-made software for data analysis applied to studying a rotational frequency of a fan. Two experiments in optics and electromagnetism advocating the use of modern digital equipment such as iOLab units or microcontrollers were presented in Ref. \cite{tufino2023}. Smartphone data was used in conjunction with an FFT data analysis in an experiment on acoustic beat phenomena in Ref. \cite{osorio2018}. The smartphone measurement app phyphox \cite{staacks2018} contains a tool for continuously viewing the FFT of measured acceleration data \cite{phyphoxweb}, and the phyphox group has listed some short example experiments on their YouTube channel \cite{phyphoxyt}.  

Jupyter notebooks are used in science education to, for example, provide interactive elements to a classroom \cite{koehler2018}, to teach basic subject-specific content while learning computational skills relevant to, e.g., data analysis or visualization \cite{weiss2017,mandanici2022}, or to report a computational activity in a form of a computational essay \cite{odden2019}. The use of Jupyter notebooks in physics education is reviewed in Ref. \cite{hanc2020} where research results supporting the use of interactive learning tools such as Jupyter notebooks are also presented.

Thanks to smartphones, more and more people carry a basic kit for vibration analysis in their pockets everywhere they go. Smartphones are increasingly utilized in physics education, and in addition to being a widely accessible measuring device, they have pedagogical potential which can be utilized in combination with traditional physics experiments \cite{girwidz2019}. Using smartphones as experimental tools can increase interest in studying physics and curiosity in the experiment itself \cite{hochberg2018}. When on-campus experiments are not possible (for example, under pandemic conditions), experiments with mobile devices provide opportunities to collect first-hand data which has been shown to lead to higher students' perceived learning success in comparison with using second-hand data \cite{klein2021}. However, another study on the secondary school level suggests that the origin of the data does not matter in achieving desired learning outcomes \cite{Priemer2020}. In the study of Ref. \cite{becker2020}, the use of mobile devices for video motion analysis reduced extraneous cognitive load during experimentation.

While the name of Fourier is usually tossed around in the context of analyzing sound frequencies, our choice to focus on vibration analysis via accelerometer data is based on making the measured quantities and measurement apparatus as simple as possible to not take attention away from the main learning objective: basic digital signal processing in data analysis. By this, we also promote the idea that there are lots of (everyday) applications for Fourier analysis beyond the conventional example of audio signals. To our knowledge, our experimental task is novel with the combination of smartphone-assisted data collection, analyzing frequencies from accelerometer data, and using a Jupyter notebook as a tool for data analysis. 

As part of this task, students practice planning an experiment within the relatively simple setup given in the task. Our experiment follows a skills-based approach to laboratory work, emphasizing the role of open inquiry and learning experimental skills such as experimental design and data analysis over cookbook-style closed instructions and reinforcing theory content. This approach has been shown to lead to beneficial outcomes both in learning critical thinking skills \cite{smith2020a,walsh2022} and in developing expert-like beliefs and attitudes about experimental physics \cite{wilcox2017,kontro2018,smith2020a,henderson2020,walsh2022}. 

This experimental task has been developed in the Erasmus+ project Developing Digital Physics Laboratory Work for Distance Learning (DigiPhysLab) \cite{lahme2022}. The produced materials include task instructions for students, additional information and suggestions for instructors, and the accompanying Jupyter notebook. All materials can be accessed on our project website \cite{digiphyslabweb}.

In Section \ref{sec:fourier} we superficially summarize the mathematical foundation for the DFT as given in the instructions of our experimental task. In Section \ref{sec:notebook} the Jupyter notebook companion is described. The experimental task and suggested possibilities for the investigation are presented along with two examples in Section \ref{sec:investigations}, and observations from our implementations of this experiment are discussed in Section \ref{sec:obs}. Our experiences and findings are summarized in Section \ref{sec:conc}. 

\section{Discrete Fourier transform} \label{sec:fourier}

The experimental task does not focus on the mathematical subtleties of the DFT. Instead, we aim to give students an intuitive understanding of what the DFT does and to introduce how the method can be applied to a signal in practice, something that can often remain vague in more mathematically oriented approaches like mathematical method courses. For completeness, a brief overview of the relevant equations is given here in the same form as it appears in our instructions for the experimental task \cite{digiphyslabweb}.

Let’s define a signal $\{x_n\}=  \left\{ x_0, x_1,\dots,x_{N-1} \right\}$ consisting of N data points (samples) taken at constant time intervals of $T_\mathrm{s}$ seconds each. The DFT of the signal is computed as 
\begin{equation}
X_k = \sum_{n=0}^{N-1} x_n e^{-\frac{i2\pi}{N}kn}, 
\end{equation}
where each $X_k$ is a complex number, and the set of $\{X_k\}=\{X_0,X_1,\dots,X_{N-1}\}$ represents the signal in the frequency domain. For a continuous function, this would correspond to a transformation of a function of time to a function of frequency. The interpretation of the Fourier-transform coefficients $X_k$ can become clearer when we look at the inverse DFT:
\begin{equation}
x_n = \frac{1}{N}\sum_{k=0}^{N-1} X_k e^{\frac{i2\pi}{N}kn}, 
\end{equation}
where we can see that the original signal can be represented as a sum of complex sinusoid components, and $X_k$ describes the amplitude and phase of each component. In this experimental task, we utilize the amplitude spectrum of the Fourier transform defined for a real-valued signal as 
\begin{equation}\label{eq:spectrum}
A_j = \frac{2}{N} |X_j|, \quad j = 0,1,\dots,N/2,
\end{equation}
which tells us how strongly the frequency $f_j$ is present in the signal. Here $\{f_j\} = \{ f_0,f_1,\dots,f_{N/2} \}$, where the frequencies are related to the sampling frequency $f_\mathrm{s} = 1/T_\mathrm{s}$ as 
\begin{equation}
f_j = \frac{kf_\mathrm{s}}{N} = \frac{j}{NT_\mathrm{s}},
\end{equation}
so that the frequencies visible in the Fourier transform never exceed $f_\mathrm{s}/2$. This maximum visible frequency is referred to as the Nyquist frequency. Note that a real-valued signal of $N$ samples in the time domain corresponds to $N/2+1$ physically meaningful frequencies in the frequency domain.

\section{The interactive notebook companion} \label{sec:notebook}

\begin{figure}
\centering  
\frame{\includegraphics[width=\textwidth]{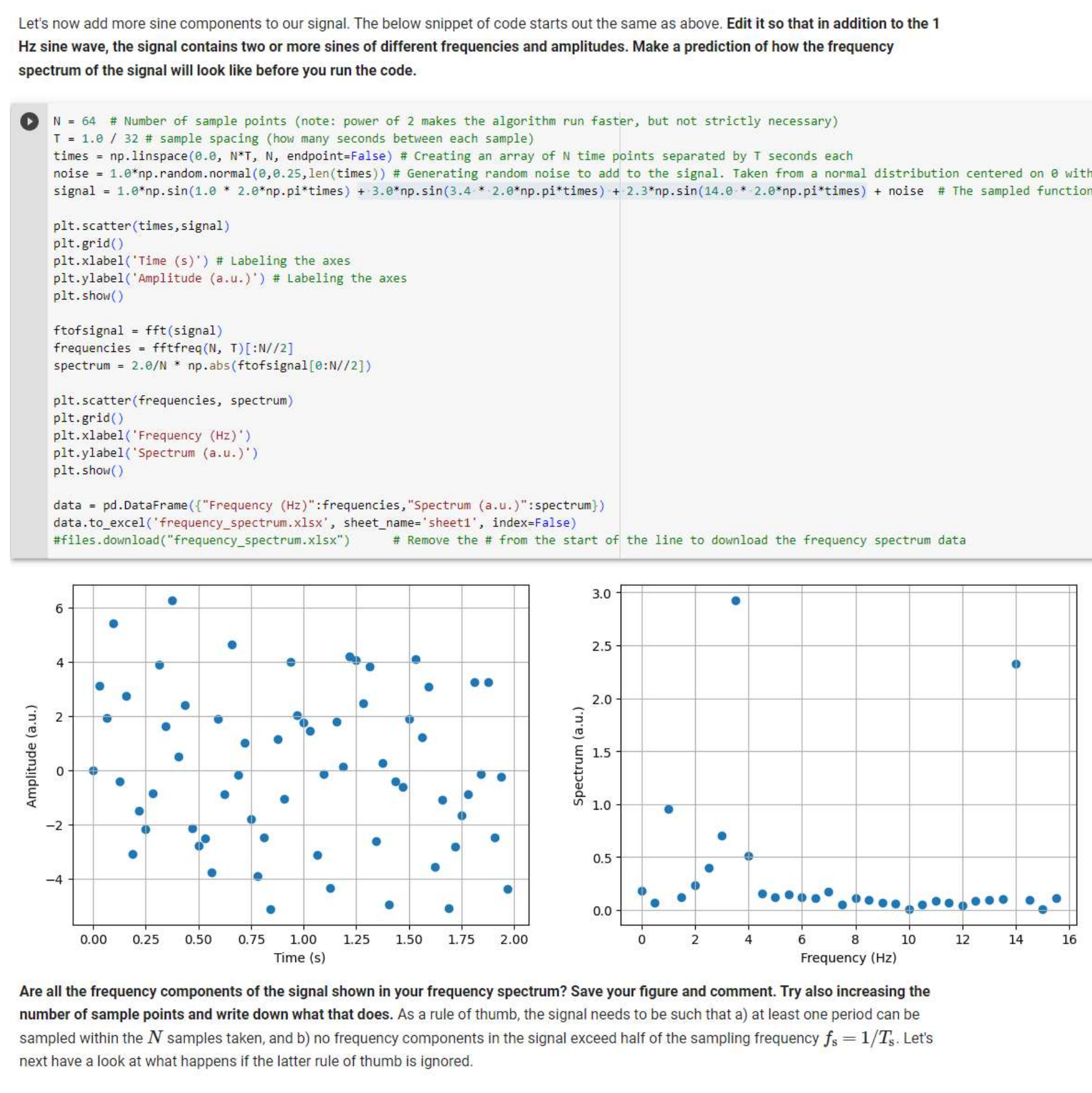}}
\caption{Snippet of the notebook companion to the experimental task. The placement of the figure output of the code was edited to make a more compact figure. Students are given small exercises in \textbf{bold} font, and the required edit to the code block is shown highlighted in pale blue.}
\label{fig:nb}
\end{figure}

To accompany the students on their journey to the basics of digital signal processing we developed a Jupyter notebook, which acts as a personal guide to the students throughout the experimental task. The notebook starts by introducing key concepts of a digital signal, such as sampling rate (sampling frequency) and signal length, via a simple sine wave example. A DFT is first applied to a simple sine function, then some random noise is added to simulate a more realistic signal, and finally, the student is asked to add a few different frequency components to the signal to see what the DFT can tell about a signal. An example snippet from the notebook is shown in Figure \ref{fig:nb}. By introducing new concepts, giving small exercises, and asking guiding questions, the notebook gently and gradually introduces the student to an intuitive understanding of what the DFT does and to the basics of Python programming as a tool for data analysis.   

%%%%%%%%%%%%%%%%%%%%%%%%%%%%%%%%%%%%%%%%%%%%%%%%%%%%%%%%%%%%%%%%%%%%%%%
%FIGURE%
%%%%%%%%%%%%%%%%%%%%%%%%%%%%%%%%%%%%%%%%%%%%%%%%%%%%%%%%%%%%%%%%%%%%%%%
%%%%%%%%%%%%%%%%%%%%%%%%%%%%%%%%%%%%%%%%%%%%%%%%%%%%%%%%%%%%%%%%%%%%%%%%

In the Jupyter notebook, we also discuss aliasing, the phenomenon of a high-frequency component of a signal masquerading as a lower frequency in a sample taken with a sampling frequency too low to accurately capture the behavior of the signal. We do not provide nor expect students to use any methods or tricks to eliminate aliasing, but we demonstrate via an example that such a phenomenon can occur and students need to be wary of it while measuring and analyzing their data. Other aspects of digital signal processing, such as windowing or noise reduction, can be added depending on the scope of the course and context but they are not necessary for the investigations of this experimental task.

Students learn about data handling when they produce a data file that can be read by the Python program for data analysis. Smartphone accelerometers used via the phyphox app typically have a maximum sampling frequency of 100 -- 400 Hz, which means that even in a short measurement there will be a lot of data points. Explicitly showing the data-handling steps in the code of the Jupyter notebook promotes methods and practices to deal with datasets with thousands of data points instead of the dozen or so often taken in educational lab experiments as a minimum requirement for meaningful statistical analysis.

Finally, the notebook provides instructions and hints for the experimental investigation of vibrations using the DFT method. Students are given the task to build their own code for data analysis by copying and modifying any parts of the examples introduced in the notebook. The measurements and analysis involved in the experimental task are described in the next section via two example investigations. 

The Jupyter notebook instruction also doubles as an example of a computational essay \cite{odden2019}, which can be used as the format of a lab report for the students' assessment. A computational essay is defined in Ref. \cite{odden2019} as \textit{"a type of essay or report that explicitly incorporates live code to support its thesis, usually written in a notebook environment"}. Therefore, a computational essay differs from a more traditional lab report in that it includes code and code output directly within the body of text, providing a natural way to describe and report a computational activity such as the data analysis of the experiment presented in this work.

\section{The experimental task}\label{sec:investigations}

Equipped with the Jupyter notebook and necessary tools for data analysis, students are then free to tackle a problem of their choice. The task for the students is to determine the frequency or frequencies present in some periodic signal they can find at home or on campus by analyzing collected smartphone accelerometer data via a discrete Fourier transform. The design of the experiment and data collection are left open for the students to decide. Some possible contexts for investigation that we have offered, or our students have come up with, include the
\begin{itemize}
\item vibrating alarm of a phone,
\item vibration of a computer (if it vibrates enough to be measured),
\item spin-dry rotation of a washing machine,
\item vibration inside a car due to the engine,
\item own heart rate,
\item vibration of an electric toothbrush,
\item person running on a treadmill,
\item any other periodic signal that students come up with.
\end{itemize}

The examples of the above list are intentionally relatively simple so that attention can be given to the main learning objective, which is the data analysis method. One can of course also use the DFT method and the notebook as a basis for investigations of more advanced physics topics later. For example, this method has been used to determine the dependence of the oscillation frequency of an elevator cabin on the length of the elevator cable after jumping in the cabin. One could also imagine examining the eigenmodes of a coupled harmonic oscillator.

\begin{figure}
\centering  
\begin{subfigure}[b]{0.49\textwidth}
\centering  
\includegraphics[width=\textwidth]{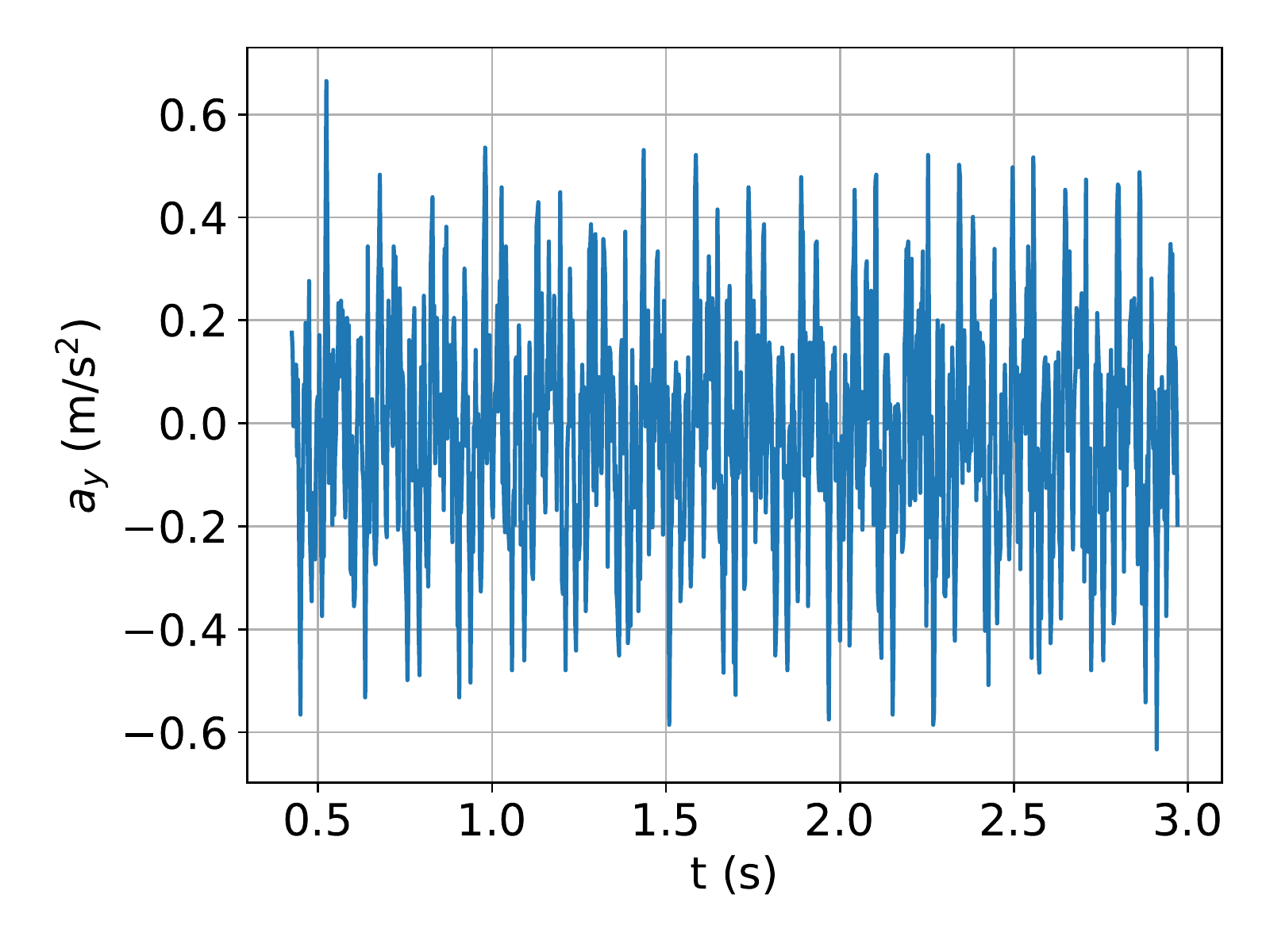}
\caption{} \label{fig:wma}
\end{subfigure}
\begin{subfigure}[b]{0.49\textwidth}
\centering  
\includegraphics[width=\textwidth]{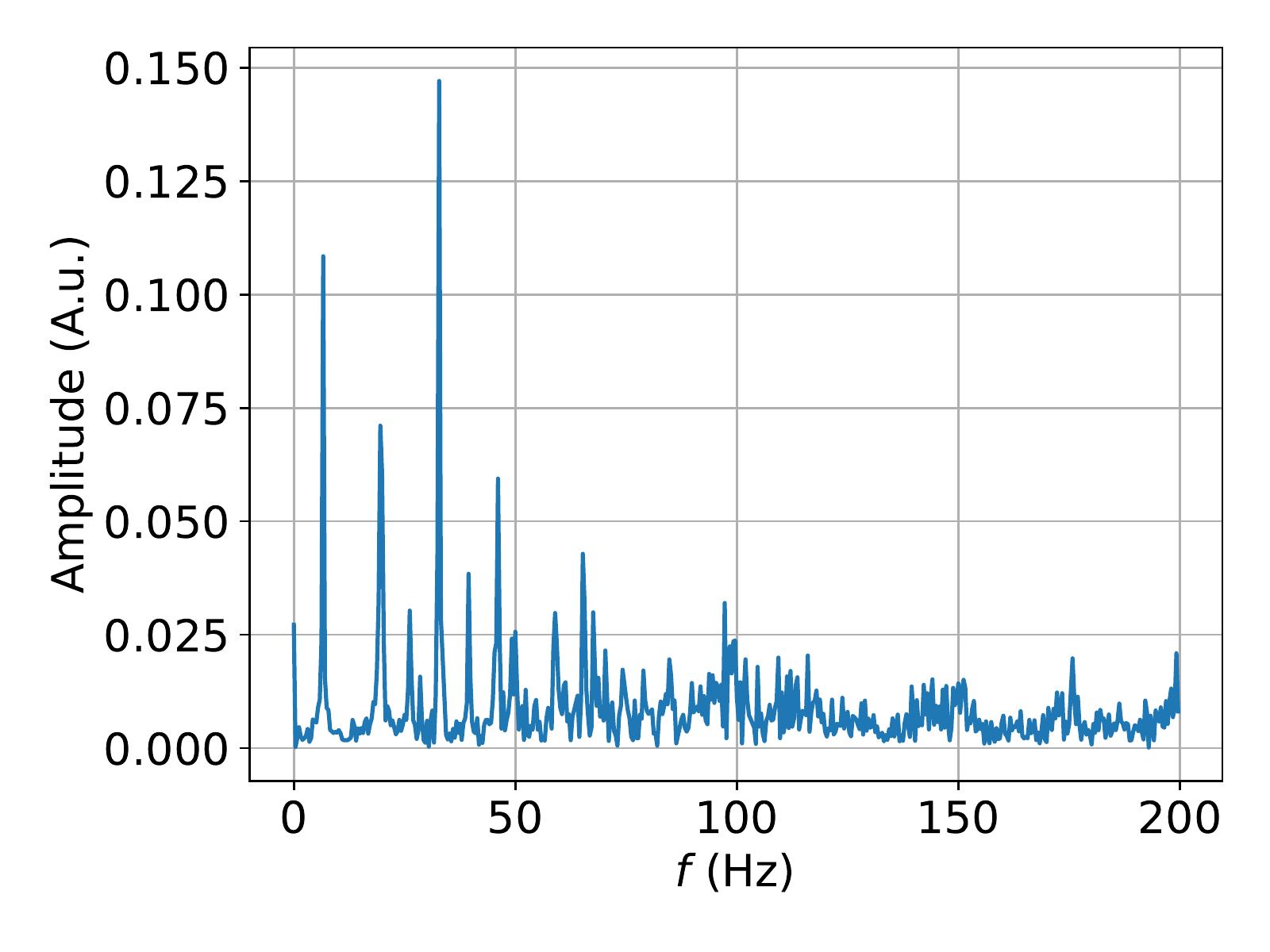}
\caption{} \label{fig:wmb}
\end{subfigure}
\caption{(a) The acceleration-vs-time signal taken with a smartphone placed on a washing machine going through a spin-dry cycle set to 400 rpm. The signal consists of 2048 samples with a sampling frequency of 400 Hz. (b) The amplitude spectrum of the signal obtained via Eq. (\ref{eq:spectrum}).}
\label{fig:wm}
\end{figure}

In the following, we highlight two example cases of the experimental task: the vibration alarm of a phone and the spin-dry sequence of a washing machine. Data was taken with the accelerometer tool of the app phyphox. In Figure \ref{fig:wma} we show data collected from the spin-dry cycle of a washing machine. A smartphone was set on top of the washing machine so that the $y$-axis of the phone's accelerometer was parallel to the most prominent movement of the washing machine. The sampling frequency was 400 Hz, and we selected a window of 2048 consecutive samples of the accelerometer data. We then performed a DFT to this selection of the data and the resulting amplitude spectrum computed via Eq. (\ref{eq:spectrum}) is shown in Figure \ref{fig:wmb}. The spectrum consists of multiple peaks at frequencies that are multiples of the first peak location at $6.6 \pm 0.4$ Hz, corresponding to $396 \pm 24$ rpm. This coincides with the 400 rpm that the spin-dry cycle of the machine was set to. The relative uncertainty of the obtained frequency is quite large. This should elicit discussions about the reliability of the DFT method for measuring small frequencies and how it can be improved.  

%%%%%%%%%%%%%%%%%%%%%%%%%%%%%%%%%%%%%%%%%%%%%%%%%%%%%%%%%%%%%%%%%%%%%%%
%FIGURE%
%%%%%%%%%%%%%%%%%%%%%%%%%%%%%%%%%%%%%%%%%%%%%%%%%%%%%%%%%%%%%%%%%%%%%%%
%%%%%%%%%%%%%%%%%%%%%%%%%%%%%%%%%%%%%%%%%%%%%%%%%%%%%%%%%%%%%%%%%%%%%%%%

In the examples presented here, we use an uncertainty estimate for the DFT-computed frequency determined by the frequency resolution $f_\mathrm{s}/N$ of the DFT, i.e., the difference between two consecutive frequency points in the amplitude spectrum rounded up to one significant figure. This is a feasible, albeit pessimistic, estimate for the uncertainty of the peak location when the peaks are sharp. For some signals, especially if the oscillation is damped, the peaks in the spectrum will be broader and it is possible to use the width parameter of a Gaussian distribution fitted to the peak of interest for an estimate of the uncertainty.  

%%%%%%%%%%%%%%%%%%%%%%%%%%%%%%%%%%%%%%%%%%%%%%%%%%%%%%%%%%%%%%%%%%%%%%%
%FIGURE%
%%%%%%%%%%%%%%%%%%%%%%%%%%%%%%%%%%%%%%%%%%%%%%%%%%%%%%%%%%%%%%%%%%%%%%%
\begin{figure}
\centering  
\begin{subfigure}[b]{0.49\textwidth}
\centering  
\includegraphics[width=\textwidth]{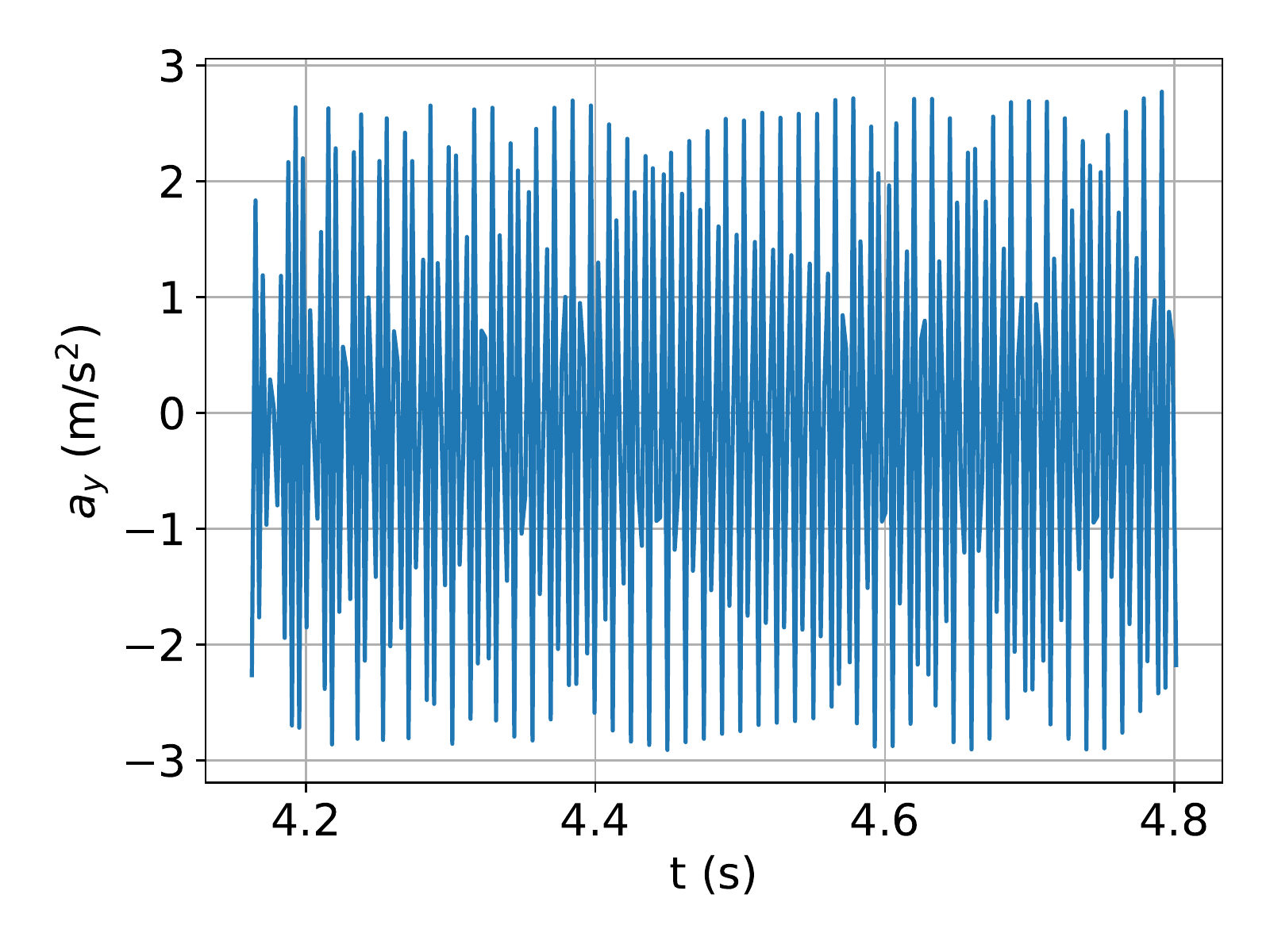}
\caption{} \label{fig:viba}
\end{subfigure}
\begin{subfigure}[b]{0.49\textwidth}
\centering  
\includegraphics[width=\textwidth]{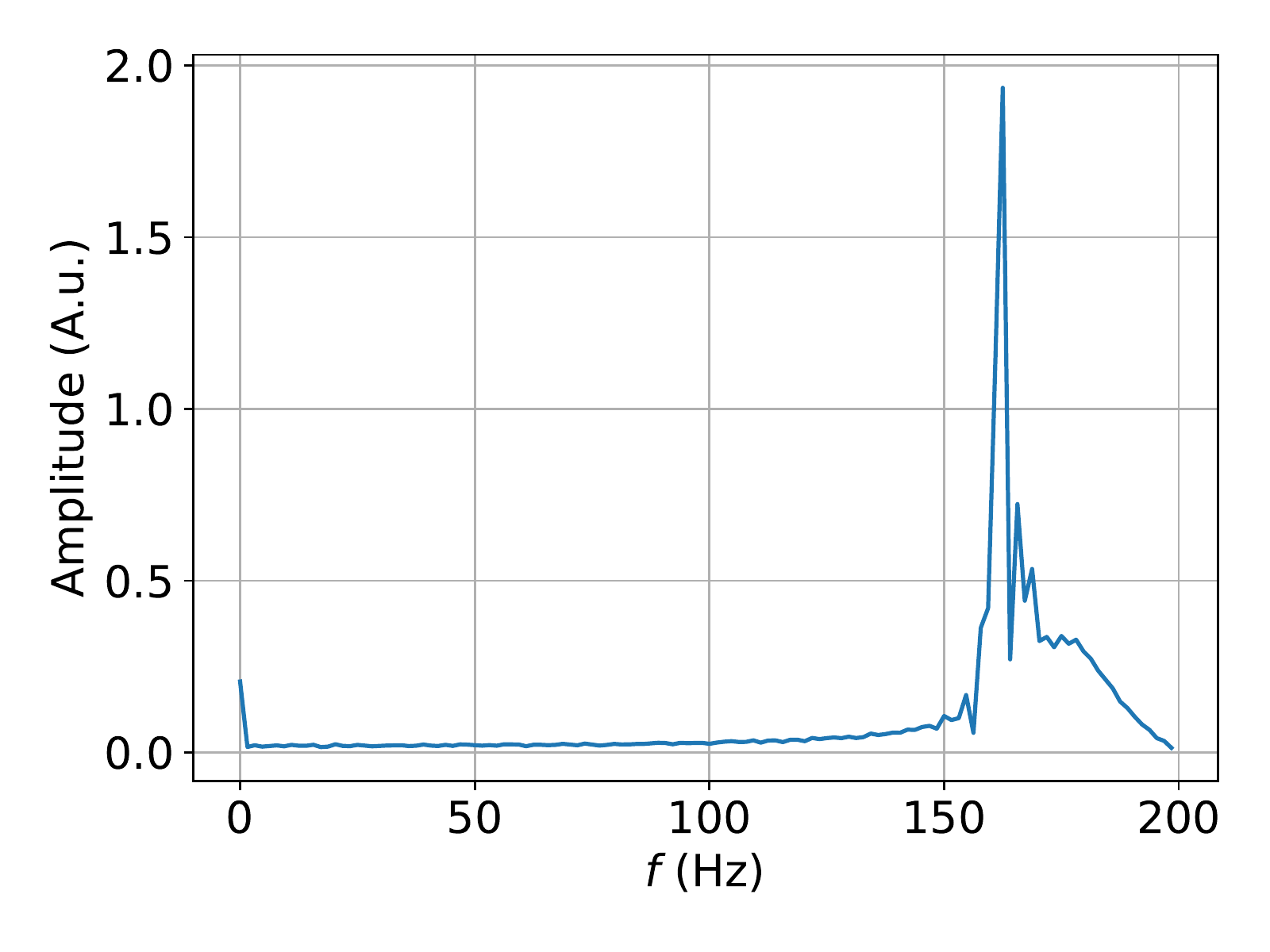}
\caption{} \label{fig:vibb}
\end{subfigure}
\caption{(a) The acceleration-vs-time signal taken with a smartphone vibrating during an incoming phone call. The signal consists of 256 samples with a sampling frequency of 400 Hz. (b) The amplitude spectrum of the signal obtained via Eq. (\ref{eq:spectrum}).}
\label{fig:vib}
\end{figure}
%%%%%%%%%%%%%%%%%%%%%%%%%%%%%%%%%%%%%%%%%%%%%%%%%%%%%%%%%%%%%%%%%%%%%%%%

As a second example, we measured the acceleration of a phone while it was lying on a table and vibrating due to an incoming call. The resulting acceleration signal is shown in Figure \ref{fig:viba}. The sampling frequency was 400 Hz and we chose a window of 256 samples where the phone was vibrating the whole time. Acceleration in the $y$ direction (parallel to the longest side) of the phone was chosen as the vibration was the most prominent in this direction. From the amplitude spectrum computed via Eq. (\ref{eq:spectrum}) and visualized in Fig. \ref{fig:vibb}, we find a single prominent peak at $162 \pm 2$ Hz. 

The typical vibration frequency for a smartphone vibration motor appears to be in the range of 130--250 Hz (based on reported numbers by manufacturers), within which the obtained frequency appears to fit very well. However, one needs to note that the accelerometer sampling frequency of 400 Hz used in this measurement is inadequate for observing frequencies higher than the Nyquist frequency of 200 Hz. Therefore, the entire range of typical vibration motor frequencies is not covered by the feasible range of our measurement, so it is entirely possible that aliasing is present in the amplitude spectrum of Fig. \ref{fig:vibb}. The possibility of aliasing can be noticed and discussed by students by comparing the observed vibration frequency to a reference value (here 130--250 Hz). As the vibration of the phone also makes an audible sound, the measured value can be checked by measuring the vibration frequency with the DFT method but using audio data where the sampling rate is typically 48 kHz. We measured the same phone vibration signal using the Audio spectrum tool in the app phyphox, which visualizes the DFT of the recorded audio signal essentially in real time. We recorded audio data on the same phone that was vibrating due to an incoming phone call. This resulted in a spectrum with a prominent peak centered around 240 Hz. Sampled with a rate of 400 Hz, the acceleration of a phone vibrating at 240 Hz appears the same as a phone vibrating at 160 Hz. This is exactly what we see in Fig. \ref{fig:vibb}, backing up the suspicion of aliasing.

The amplitude spectra of Figures \ref{fig:wmb} and \ref{fig:vibb} are fundamentally different: for the smartphone vibration we have a single prominent peak while for the washing machine, there are several peaks at frequencies that are multiples of the basic frequency of the lowest peak. The vibration of the smartphone appears simpler than that of the washing machine in terms of sinusoidal components, as the smartphone vibration is characterized by essentially a single sinusoid (although there is a low-amplitude tail attached to the most prominent peak in the spectrum, and some details of the spectrum could be lost due to aliasing). The shaking of a washing machine during its spin-dry cycle requires several frequency components that conspire to cancel and amplify each other suitably to describe the observed movement.

\section{Observations and students' perception of the task}\label{sec:obs}

We have observed that enabling students to choose their own target for investigation creates engagement and excitement as the outcome of the experiment is not known to anyone beforehand and students typically choose a target that they are genuinely interested in learning more about. This is supported by the findings of Refs. \cite{kalender2021,schmidt2018} that freedom and choice afforded by open-ended lab work can be beneficial for student engagement and enjoyment.

Based on our experience with running this task, students are in general able to connect the peaks of the computed amplitude spectrum to the physical system that they chose to investigate and they can make meaningful conclusions based on their measured data. Therefore, it appears that the main learning objective of applying the DFT in authentic digital signal processing is typically fulfilled.

A common pitfall for students involves the definition of the signal length $N$ and sample spacing $T_\mathrm{s}$ when moving from artificial examples to real measurement. The length of the signal is the number of samples in the data window chosen for examination. The sampling rate (and thus the sample spacing) is determined by the specifics of the accelerometer in the smartphone together with the measuring app. Despite a direct prompt in the notebook instructions to carefully identify the sample spacing and signal length from the data, this is a point that has most often required help from an instructor. Another point where an instructor is often needed is to remind the student to crop the accelerometer data to a time window where only the examined signal is present so that any artifacts from starting and ending the measurement or periods when the signal is not present are not taken into the analysis. We noted that these two points might be overlooked without an instructor present in the data analysis part of the task. Therefore, prompts to take them into account have been strengthened in the current version of the notebook instructions. 

In our latest implementation of this experimental task at the University of Jyväskylä, we collected data using a questionnaire \cite{lahme2023} developed in the DigiPhysLab project for evaluating the quality of an individual lab task. The task was administered as a practical exercise on a graduate-level mathematical method course on integral transforms, featuring also the DFT. The participating students were physics majors on average in their 4th year of university physics studies. We received four responses to our questionnaire. Students rated the use of digital technology for data collection and analysis as positive in this experimental task. The task instructions were perceived as adequate and clear, though one must bear in mind that the participating students had encountered Python programming before. Students reported that the task was interesting, easy to understand, and a good demonstration of how the mathematical method can be used in practice.

Based on the students' feedback on the task, we improved the way data input was done in the Jupyter notebook. Data was previously handled in a rudimentary way using very simple two-column text files due to a desire to make the data input work easily regardless of which notebook environment was used by the student. This was somewhat confusing, though, as measurement data from the smartphone accelerometer is typically exported as an Excel or CSV file. Moreover, we figured it was not promoting very good data-handling practices to require cropping the data into a text file by hand. Therefore, the data input was changed so that the files produced by the measuring app can be read directly, and cropping of the data is done in the code in the notebook.  

\section{Conclusions}\label{sec:conc}
We have presented an experimental task for university physics involving concepts of digital signal processing in an everyday context. Especially highlighted is the discrete Fourier transform (DFT), which is introduced in a highly practical way via an interactive Jupyter notebook and used to perform vibration analysis on a target chosen by the student. The task can be conducted both in distance-learning and on-campus scenarios, as suitable sources of vibration can be found in most places.   

In our experience, the experimental task works as a way to teach about data analysis, especially the basics of digital signal processing. After completing the task students are able to apply a DFT to authentic data and make meaningful conclusions based on the analysis of the amplitude spectrum. In general, students have also found this task interesting and engaging. 

This experimental task can be implemented as a full laboratory exercise, with a focus on experimental skills such as data analysis and estimating measurement uncertainty. For such implementations, a computational essay is suggested as the lab report format for assessment. The task can also be used as a lighter exercise or a demonstration to give a practical connection in, for example, more theoretically oriented method courses. Instead of a full lab report one can then, for example, administer group discussions about the obtained results, or assign a more concise report with only the most essential aspects of the measurements and results. While the Jupyter notebook used for data analysis provides insight into how the DFT is applied to data, some ideas of the experimental task can be conveyed as a demonstration also by using the Acceleration Spectrum tool in the app phyphox (or similar), which continuously computes and visualizes the DFT of the accelerometer data.    

This task serves both theoretically and experimentally oriented students with its hands-on computational approach to a subject often discussed in a highly abstract and theoretical manner. The experimental task in its entirety along with more background information to instructors is available at the DigiPhysLab project website \url{https://jyu.fi/digiphyslab}.

\section*{Acknowledgments}

This work was supported by the Erasmus+ 2020-1-FI01-KA226-HE-092531 project Developing Digital Physics Laboratory Work for Distance Learning -- DigiPhysLab.

\section*{Data availability statement}
The data that support the findings of this study are available upon reasonable request from the authors.

\section*{Ethical statement}
As per the research ethics requirements of the University of Jyväskylä the research did not require approval from the ethics committee. The research was conducted in accordance with the principles embodied in the Declaration of Helsinki and in accordance with local statutory requirements. All participants gave written informed consent to participate in the study.

\section*{References}
\bibliography{mybibfile}

\end{document}